\documentclass[11pt,a4paper]{article}
\usepackage{amsmath}
\usepackage[psamsfonts]{amssymb}
\usepackage{amsmath}
\usepackage{epsfig}
\usepackage{graphicx}

\title{Cosmological entropy and generalized second law of
thermodynamics in $F(R,G)$ theory of gravity}
\author{H. Mohseni Sadjadi\footnote{mohsenisad@ut.ac.ir},
\\{\small Department of Physics, University of Tehran,}
\\{\small P.O.B. 14395-547, Tehran, Iran}.}

\begin{document}
\maketitle

\begin{abstract}
We consider a spatially flat Friedmann-Lemaitre-Robertson-Walker
space time and investigate the second law and the generalized
second law of thermodynamics for apparent horizon in generalized
modified Gauss Bonnet theory of gravity (whose action contains a
general function of Gauss Bonnet invariant and the Ricci scalar:
$F(R,G)$). By assuming that the apparent horizon is in thermal
equilibrium with the matter inside it,  conditions which must be
satisfied by $F(R,G)$ are derived and elucidated through two
examples: a quasi-de Sitter space-time and a universe with power
law expansion.
\end{abstract}

\section{Introduction}
To explain the present accelerated expansion of the universe,
which is confirmed by several astrophysical data \cite{acc},
different models have been introduced. One of these models, dubbed
as dark energy model, assumes that the universe is dominated by an
exotic matter with negative pressure. This kind of matter violates
the strong energy condition $P>-{\rho\over 3}$. Besides, models in
which the Einstein theory of gravity is modified, have been also
used to describe the present acceleration of the universe
\cite{infmod}. In these models there is no need for exotic matter
with odd properties, but instead, the action contains general
function of invariants obtained from the Riemann curvature tensor
such as the Ricci scalar, $R$, or Gauss-Bonnet invariant term, $G$
(such a model appears also in the low energy limit of heterotic
string theory \cite{string}). Consistency with other laws of
physics (e.g. thermodynamics laws considered in this paper) and
astrophysical data put many conditions on these modified theories.

Thermodynamics aspects of general relativity and cosmology, such
as validity of thermodynamics laws for cosmological
horizons\cite{bousso}, thermodynamics of dark energy and so on
have also been the subject of many studies in the past and recent
years. The generalized second law (GSL) of thermodynamics in
Einstein theory of gravity was investigated in \cite{dav}, and was
extended to the case of dark energy in \cite{pav}. The study of
this law for the {\it future event horizon} in the simple modified
theory of gravity whose action contains only a function of $R$ or
$G$, can be found in \cite{sad}.

Recently, the generalized modified Gauss-Bonnet gravity, whose the
action contains a general function of $R$ and $G$ ($F(R,G)$), has
attracted more attention. Besides its stability, this is due to
its ability to describe the present acceleration of the universe
as well as the phantom divide line crossing and transition from
acceleration to deceleration phases \cite{al}.

The scheme of the paper is as follows:

We consider the generalized modified Gauss-Bonnet theory of
gravity in a spatially flat Friedman-Robertson-Walker (FRW) space
time. As in this theory there may be many choices for $F(R,G)$,
all leading to a same dynamics, GSL by putting some constraints on
$F(R,G)$ can be considered as a test to choose a viable model. We
study the thermodynamics second law and generalized second law of
thermodynamics for the {\it apparent horizon} (there are many
studies about the thermodynamics aspects of the apparent horizon,
for a review see \cite{app}). We will assume that the apparent
horizon is in thermal equilibrium with the matter. We investigate
the conditions obeyed by $F(R,G)$ to satisfy these thermodynamics
laws and show that some of these conditions are in agreement with
the other constraints required for dynamical stability of the
model discussed before in the literature.  We elucidate our
results through two important cosmological examples: the quasi-de
Sitter space-time which is a good candidate for the early stage
and late time evolution of the universe, and a space-time with
power law expansion filled by ordinary matter such as dust and
radiation.

Units with $\hbar=c=G_N=k_B=1$ are used in the paper.

\section{ Apparent horizon entropy and cosmological GSL in $F(R,G)$ model}
\subsection{preliminaries}
$F(R,G)$ model of gravity is described by the action
\begin{equation}\label{1}
S=\int \left({1\over
16\pi}F(R,G)+\mathcal{L}_m\right)\sqrt{-g}d^4x,
\end{equation}
where $\mathcal{L}_m$ is the matter lagrangian density and
$F(R,G)$ is a function of the Ricci scalar curvature and
Gauss-Bonnet invariant defined by
\begin{equation}\label{2}
G=R^2-4R_{\mu \nu}R^{\mu \nu}+R_{\mu \nu \rho \sigma}R^{\mu \nu
\rho \sigma}.
\end{equation}
By varying the action with respect to the metric components
$g_{\mu \nu}$, one obtains the (gravitational) field equations:
\begin{eqnarray}\label{3}
&&8\pi T^{\mu \nu}=g^{\mu
\nu}\nabla^2F_R-\nabla^{\mu}\nabla^{\nu}F_R-{1\over 2}g^{\mu\nu}F
+R^{\mu\nu}F_R
+2g^{\mu \nu} R\nabla^2 F_G\nonumber \\
&&-4R^{\mu \nu}\nabla^2 F_G -2R(\nabla^{\mu}\nabla^{\nu})F_G
+4R^{\mu \alpha}\nabla_{\alpha}\nabla^{\nu}F_G+4R^{\nu
\alpha}\nabla_{\alpha}\nabla^{\mu}F_G
\nonumber \\
&&-4g^{\mu \nu}R^{\alpha \beta}\nabla_{\alpha}\nabla_{\beta}F_G
+4R^{\mu \alpha\nu\beta}\nabla_{\alpha}\nabla_{\beta}F_G+2RR^{\mu
\nu}F_G-4R^{\mu}_{\alpha}R^{\nu \beta}F_G
\nonumber\\
&&+2R^{\mu \alpha\beta\gamma}R^{\nu}_{\alpha\beta\gamma}F_G
+4R_{\alpha \beta}R^{\mu\alpha\beta\nu}F_G,
\end{eqnarray}
where $F_R={\partial F(R,G)\over \partial R},\,\,\,F_G={\partial
F(R,G)\over \partial G}$ and so on, and $T^{\mu \nu}$ is the
energy-momentum tensor of matter. We consider a spatially flat
Friedmann-Robertson-Walker (FRW) space time in comoving
coordinates
\begin{equation}\label{4}
ds^2=-dt^2+a^2(t)(dx^2+dy^2+dz^2),
\end{equation}
where $a(t)$ is the scale factor. In terms of the Hubble parameter
$H={\dot{a}\over a}$, where the over dot indicates derivative with
respect to the time $t$, we have
\begin{equation}\label{5}
R=6(\dot{H}+2H^2),\,\,\,G=24H^2(\dot{H}+H^2).
\end{equation}
Equation (\ref{3}) gives the following equations
\begin{eqnarray}\label{6}
F_R\dot{H}&=&-4\pi(P_m+\rho_m)+{1\over
2}(H\dot{F_R}-\ddot{F_R}+4H^3\dot{F_G}-8H\dot{H}\dot{F_G}\nonumber \\
&&-4H^2\ddot{F_G}),\nonumber\\
F_R H^2&=&{8\pi\over 3}\rho+{1\over
6}(F_RR-F-6H\dot{F_R}+GF_G-24H^3\dot{F_G}).
\end{eqnarray}
$\rho_m$ and $P_m$ are energy density and pressure of the matter
component which behaves as a perfect fluid at large scale,
satisfying the continuity equation
\begin{equation}\label{7}
\dot{\rho_m}+3H(P_m+\rho_m)=0.
\end{equation}
In the above, the matter ingredient may consist of various
interacting components.
\subsection{Apparent horizon entropy}
In terms of the Hubble parameter, the apparent horizon radius is
given by $R_h={1\over H}$. The entropy of this dynamical horizon,
$S_h$, obtained by the Noether charge method \cite{ent} is given
by
\begin{equation}\label{8}
S_h=-{1\over 8}{\int}_{Horizon}\left(F_R{\partial R\over\partial
R_{\alpha \beta \gamma \rho}}+F_G{\partial G\over\partial
R_{\alpha\beta\gamma\rho}}\right)\varepsilon_{\alpha
\beta}\varepsilon_{\gamma \rho}dA_{h} ,
\end{equation}
where $\varepsilon_{\mu \nu}$ are binormal vectors to horizon
surface and $dA_h$ is the differential surface element. Applying
this result to the apparent horizon in the FRW space time yields:
\begin{equation}\label{9}
S_h=\pi \left(H^{-2}F_R+4F_G\right).
\end{equation}
Dynamics of the horizon entropy is described by
\begin{equation}\label{10}
\dot{S_h}=-2\pi H^{-3}\dot{H}F_R+\pi H^{-2}\dot{F}_R+4\pi
\dot{F}_G.
\end{equation}
Note that a linear term in Gauss-Bonnet invariant in $F$, may
change the horizon entropy but has no influence on its time
derivative.

For a de Sitter space-time, $\dot{H}=0$, and therefore
$\dot{S_h}=0$. If $F(R,G)=R$, for $\dot{H}<(>)0$, we have
$\dot{S_h}>(<)0$. So in a super-accelerated universe the second
law of thermodynamics does not hold for the apparent horizon in
Einstein theory of gravity. This may not be true in modified
theories of gravity. To show this, let us consider a quasi-de
Sitter space-time which depends mildly on time:
\begin{equation}\label{11}
H=H_0+H_0^2\epsilon t+\mathcal{O}(\epsilon^2);\hspace{3mm}
\epsilon={\dot{H}\over H^2};\,\,\dot{\epsilon}
=\mathcal{O}(\epsilon^2),
\end{equation}
where $\mid\epsilon\mid\ll 1$. By expanding $R$ and $G$ around de
Sitter point, $H_0$, up to order $\mathcal{O}(\epsilon^2)$, as
\begin{eqnarray}\label{12}
R&=&6H^2(2+\epsilon)=12H_0^2+6H_0^2(1+4H_0t)\epsilon+\mathcal{O}(\epsilon^2)\nonumber  \\
G&=&24H^4(1+\epsilon)=24H_0^4+24H_0^4(1+4H_0t)\epsilon+\mathcal{O}(\epsilon^2),
\end{eqnarray}
and by using
\begin{eqnarray}\label{13}
&&\dot{F_R}=(F_{RR}(H_0)+F_{RRR}(H_0)(R-R_0)+F_{RRG}(H_0)(G-G_0))\dot{R}\nonumber\\
&&+
(F_{RG}(H_0)+F_{RRG}(H_0)(R-R_0)+F_{RGG}(H_0)(G-G_0))\dot{G},\nonumber\\
&&\dot{F_G}=(F_{GR}(H_0)+F_{GRR}(H_0)(R-R_0)
+F_{RGG}(H_0)(G-G_0))\dot{R}\nonumber \\
&&+ (F_{GG}(H_0)+F_{RGG}(H_0)(R-R_0)+F_{GGG}(H_0)(G-G_0))\dot{G},
\end{eqnarray}
where $G_0=G(H_0)$ and $R_0=R(H_0)$, we conclude that the
thermodynamics second law for the horizon, $\dot{S_h}\geq 0$,
implies
\begin{equation}\label{14}
\left( 192 F_{GG}(H_0) H_0^6+
96F_{RG}(H_0)H_0^4+12F_{RR}(H_0)H_0^2-F_R(H_0)\right) \epsilon\geq
0.
\end{equation}
Equation (\ref{6}) results in that in a de Sitter space-time
$P_m+\rho_m=0$. Hence from the continuity equation we find out
that $\rho$ is a constant: $\rho_m(H_0)=\Lambda\in \Re^{+}$. The
solution with $\rho_m(H_0)=0$ is dubbed as de Sitter vacuum
solution. In general, $F$ can be any function that satisfies
\begin{equation}\label{15}
16\pi \rho_m(H_0)+6H_0^2F_R(H_0)+24H_0^4F_G(H_0)-F(H_0)=0
\end{equation}
provided that (\ref{15}) has a positive root $H_0>0$.

In contrast to the Einstein theory of gravity, depending on the
form of $F$, (\ref{14}) may be respected. To see this Let use
choose a model with $F_R(H_0)<0$. For the vacuum de Sitter
solution ($\rho_m(H_0)\approx 0$), the stability of $F(R,G)$ model
requires \cite{stab}
\begin{equation}\label{16}
{F_R(H_0)\over
12H_0^2F_{RR}(H_0)+96H_0^4F_{RG}(H_0)+192H_0^6F_{GG}(H_0)}>1.
\end{equation}
By comparing (\ref{14}) and (\ref{16}), we find out that in a
stable model, thermodynamics second law for the apparent horizon
holds whenever $\dot{H}> 0$ .

\subsection{GSL}
Now let us consider time evolution of entropy of the matter inside
the horizon, denoted by $S_{in}$, and study the GSL which states
that the sum of the horizon entropy and the matter entropy is not
decreasing in time:
\begin{equation}\label{17}
\dot{S}_{tot}=\dot{S}_{in}+\dot{S}_h\geq 0.
\end{equation}
In this way the horizon entropy is related to the information
behind it. Hence we can consider the entropy of the universe as
the sum of the entropy of the matter inside the horizon, and the
horizon entropy.

In the absence of matter, or when the role of matter is
negligible, GSL reduces to thermodynamics second law for the
apparent horizon discussed before. From the first law of
thermodynamics:
\begin{equation}\label{18}
dE=TdS_{in}-PdV,
\end{equation}
where $V={4\pi\over 3H^3}$, and the continuity equation, we obtain
\begin{equation}\label{19}
\dot{S}_{in}=-8\pi^2 H^{-5}(\dot{H}+H^2)(P_m+\rho_m).
\end{equation}
The matter and the horizon are in thermal equilibrium and, in
analogy with black hole thermodynamics, we have taken the horizon
temperature $T={1\over 2\pi H}$. For an accelerated expansion
$(\dot{H}+H^2)>0$ and for ordinary matter whose the pressure is
positive, we have $\dot{S}_{in}< 0$. So a necessary condition for
the GSL to hold is $\dot{S}_h>0$.

In the case that the universe is dominated by a barotropic perfect
fluid (at large scale), $P_m=w_m\rho_m$, the continuity equation
yields
\begin{equation}\label{20}
\rho=\tilde{\rho_0}a^{-3\gamma},
\end{equation}
and the generalized second law can be written as
\begin{equation}\label{21}
-2\pi H^{-3}\dot{H}F_R+\pi H^{-2}\dot{F}_R+4\pi
\dot{F}_G-8\pi^2H^{-5}(H^2+\dot{H})\gamma
\tilde{\rho_0}a^{-3\gamma}\geq 0.
\end{equation}
We have defined $\gamma=w_m+1$. But in more general cases the
matter component may consist of several barotropic fluids and
$w_m$ is not necessarily a constant, and obtaining an analytical
solution for matter density is not straightforward. In these cases
it is more convenient to use (\ref{6}), and write (\ref{19}) as
\begin{equation}\label{22}
\dot{S}_{in}=\pi H^{-5}(\dot{H}+H^2)
(2F_R\dot{H}-H\dot{F_R}+\ddot{F_R}-4H^3\dot{F_G}+8H\dot{H}\dot{F_G}
+4H^2\ddot{F_G}).
\end{equation}
Hence GSL reads
\begin{eqnarray}\label{23}
\dot{S}_{tot}&=&2\pi H^{-5}\dot{H}^2F_R-\pi
H^{-4}\dot{H}\dot{F_R}+4\pi H^{-4}\dot{H}(H^2+2\dot{H})\dot{F_G}
\nonumber \\
&&+4\pi H^{-3}(H^2+\dot{H})\ddot{F_G}+\pi
H^{-5}(\dot{H}+H^2)\ddot{F_R}\geq 0.
\end{eqnarray}

For a de Sitter space-time the universe undergoes an adiabatic
reversible expansion. Although the horizon entropy is not
necessarily increasing in Einstein theory of gravity, but GSL
holds generally in this theory: $\dot{S}_{tot}=2\pi
H^{-5}{\dot{H}}^2\geq 0$.

To elucidate the r\^{o}le of matter entropy in GSL, let us
reconsider the quasi-de Sitter space-time. As an interesting
result the expression linear in $\epsilon$ in $\dot{S}_h$ is
cancelled out with the corresponding expression in $\dot{S}_{in}$,
and $\dot{S}_{tot}$ when expanded in terms of $\epsilon$,
(depending on the model) begins with $\epsilon^2$ or
$\dot{\epsilon}$. To be more specific and to elucidate more
explicitly our results, let us study the conditions that GSL puts
on $F$ in a model whose the Hubble parameter is given by:
\begin{equation}\label{24}
H=H_0+{H_1\over t},\,\,\,H_1>0.
\end{equation}
This universe tends to (quasi-) de Sitter space-time at late time,
i.e. when $t^2\gg {\mid H_1 \mid \over H_0^2}$. We do not restrict
ourselves to one component barotropic matter. By using
\begin{eqnarray}\label{25}
&&\ddot{F_R}=F_{RR}\ddot{R}+F_{RG}\ddot{G}+F_{RRR}\dot{R}^2+2F_{RRG}\dot{R}\dot{G}+
F_{RGG}\dot{G}^2\nonumber
\\
&&\ddot{F_G}=F_{RG}\ddot{R}+F_{GG}\ddot{G}+F_{RRG}\dot{R}^2+2F_{RGG}\dot{R}\dot{G}
+F_{GGG}\dot{G}^2,
\end{eqnarray}
and expanding $F_{R}$, $F_G$, $F_{RRR}$, and son on around the de
Sitter point as what was done in (\ref{13}), after some
computations we obtain
\begin{eqnarray}\label{26}
&&\dot{S_h}=-{2\pi H_1\over
H_0^2t^2}\left(12H_0F_{RR}+96H_0^3F_{RG}+192H_0^5F_{GG}-{1\over
H_0}F_R\right)\nonumber \\
&+&\mathcal{O}\left({1\over t^3}\right),
\end{eqnarray}
and,
\begin{eqnarray}\label{27}
&&\dot{S}_{tot}={48\pi H_1\over H_0^2}
\left(8H_0^2F_{RG}^2(H_0)+F_{RR}(H_0)+16H_0^4F_{GG}(H_0)\right){1\over t^3}\nonumber \\
&&+\mathcal{O}\left({1\over t^4}\right).
\end{eqnarray}

Therefore GSL is satisfied provided that
\begin{equation}\label{28}
8H_0^2F_{RG}^2(H_0)+F_{RR}(H_0)+16H_0^4F_{GG}(H_0)>0.
\end{equation}

Note that this result is general, in the sense that it is
independent of the explicit form of  $F(R,G)$. To go further, let
us consider the specific functional form for $F(R,G)$ suggested in
the modified Gauss-Bonnet gravity: $F(R,G)=R+g(G)$ where $g$ is a
function of $G$. Viable models are specified by the following
conditions \cite{stab}: $g(G)$ and its derivatives with respect to
$G$ are continuous; $g_{GG}>0$ for $G<G_0$, where $G_0=G(H_0)$ ,
and $\lim_{G\to \infty}g_{GG}=0$, and finally
$0<H_0^6g_{GG}(H_0)<1/384$. Also stability of the model requires
that $g_{GG}>0$ for $G\leq 24H_0^4$. Therefore in this model
(\ref{28}) is satisfied. {\it It is worthwhile noting that the
validity of GSL, in contrast to validity of thermodynamics second
law discussed after (\ref{16}), requires that at early times the
universe must be in quintessence phase $\dot{H}<0$}.

For Einstein theory of gravity we have
\begin{equation}\label{29}
\dot{S}_{tot}={2H_1^2\over H_0^5}{1\over t^4}+
\mathcal{O}\left({1\over t^5}\right),
\end{equation}
and although the thermodynamics second law is only true for
$H_1>0$, the GSL is valid generally at late time in this model as
was expected.

At the end by considering the power law expansion in modified
Gauss-Bonnet gravity, we examine the validity of GSL and its
consequence on the Hubble parameter. The importance of this
example lies on the fact that we can obtain an explicit expression
for $F(R,G)$. T
he accelerated power law expansion of the FRW
universe is described by
\begin{equation}\label{30}
a=a_0t^m,\hspace{3mm} m>1,\hspace{3mm}   t\geq 0.
\end{equation}
We also assume that the universe is dominated by a barotropic
matter whose equation of state is given by $P_m=w_m\rho_m$. In
this model as we use modified gravity there is no needs to employ
non-ordinary matter.

The Hubble parameter is obtained as $H={m\over t}$, and the
continuity equation yields
\begin{equation}\label{31}
\rho=\rho_0t^{-3m\gamma},
\end{equation}
where $\gamma=w_m+1$. We consider a solution to (\ref{6}) in the
form $F=R+g(G)$, where $g$ satisfies the following equation
\begin{equation}\label{32}
{4\over m-1}\tilde{G}^2g_{\tilde{G}\tilde{G}}+
\tilde{G}g_{\tilde{G}}-g=6m^2\tilde{G}^{1\over 2} -K
\tilde{G}^{{3m\gamma\over 4}}.
\end{equation}
We have defined $\tilde{G}:={G\over 24m^3(m-1)}=t^{-4}$ and
$K=16\pi \rho_0$. The solution to the above equation is
\begin{equation}\label{33}
g=C_1 \tilde{G}+C_2\tilde{G}^{1-m\over 4}-{4K(m-1)\over
(-4+3m\gamma)(-1+(1+3\gamma)m)} \tilde{G}^{3m\gamma\over
4}-{12m^2(m-1)\over (1+m)}\tilde{G}^{1\over 2}
\end{equation}
Nor the field equation neither the thermodynamics second law is
affected by the linear term in Gauss-Bonnet invariant, so we set
$C_1=0$. Authors In \cite{end} argued that for $K=6m^2$ and
$m\gamma=2/3$ we must have $g=0$ and therefore $C_2=0$. Their
argument is based on the fact that in the Einstein theory of
gravity we have $K=6m^2$ and $m\gamma=2/3$, but we must note that
the converse is not true. So although $C_2=0$ is a possible
choice, it is not necessary, and one can take $C_2\neq 0$. From
(\ref{23}) one can verify that the GSL is valid when
\begin{equation}\label{34}
-{1\over 3m^3(m-1)}\tilde{G}^{5\over 4}g_{\tilde{G}\tilde{G}}
+{1\over m^2}\tilde{G}^{-{1\over 4}} -{K\gamma\over
4}\left({m-1\over m^4}\right)\tilde{G}^{{3\over 4}(m\gamma-1)}\geq
0.
\end{equation}
By putting (\ref{33}) in (\ref{34}), we conclude that GSL holds
only when
\begin{equation}\label{35}
\tilde{C_2}t^{m+1}+{\gamma K\over 4}\left({1\over
1-(1+3\gamma)m}+{m-1\over m^2}\right)t^{2-3m\gamma}\leq {1\over
m+1},
\end{equation}
where $\tilde{C_2}={(3+m)C_2\over 48m}$. It is clear that an
adiabatic expansion is not possible in this model, hence the
equality must be excluded in the above relation.

The only free parameter in $F(R,G)$ is $C_2$.  (\ref{35}) implies
that GSL is always satisfied for $C_2=0$ and a FRW universe
characterized by
\begin{equation}\label{1000}
0<m<{2+3\gamma-\sqrt{4+9\gamma^2}\over 6\gamma},
\end{equation}
or
\begin{equation}\label{1001}
{1\over 1+3\gamma}<m<{2+3\gamma+\sqrt{4+9\gamma^2}\over 6\gamma}.
\end{equation}

So the solution proposed in \cite{end} is in agreement with GSL
provided $m$ specified by (\ref{30}) satisfies (\ref{1000}) or
(\ref{1001}). Indeed GSL puts some conditions on the scale factor
and therefore on the evolution of the universe.

But it seems that we cannot save this law  generally in our
method. For example in a model with
$m>{2+3\gamma+\sqrt{4+9\gamma^2}\over 6\gamma}$ and for ordinary
matter $\gamma\geq 1$, GSL does not hold in the limit $t\to 0$(
more precisely $Kt^{2-3m\gamma}\gg G_N$), or large curvature
limit. This may be due to the fact that in our classical
computation we have ignored the role of quantum gravity and
quantum effects which become important in the large curvature
limit (or when $t\to 0$) or large energy densities.
\section{conclusion}
In this paper we studied the second and generalized second laws
(GSL) of thermodynamics for the apparent horizon in a spatially
flat FRW universe in the framework of generalized modified Gauss
Bonnet theory of gravity (whose action contains a general function
of Gauss Bonnet invariant and the Ricci scalar: $F(R,G)$). We
computed the horizon entropy via Noether charge method (see
(\ref{9})), and the matter entropy via Friedmann equations to
obtain general expressions for the total entropy and its time
derivative. We assumed that the horizon temperature which is given
by Gibbons-Hawking temperature is the same as the temperature of
the matter inside the horizon. It was shown that in the Einstein
theory of relativity, the apparent horizon entropy decreases in
super-accelerated universe, which may not be the case in $F(R,G)$
theory of gravity. This fact was shown through an example in the
"Apparent-horizon entropy" subsection. It must be noted that in
the absence of matter or when the contribution of matter entropy
is negligible, GSL  reduces to the thermodynamics second law for
the apparent horizon. It was shown that in an accelerated
expanding universe filled with ordinary matter, the matter entropy
decreases with time (see (\ref{19})) so we must take also into
account the contribution of the horizon entropy to get a total
entropy which increases with time satisfying GSL. Although in
$F(R,G)$ theory of gravity, there may be many choices for $F(R,G)$
which satisfy Friedmann equations and all lead to a same dynamics
it was shown that GSL by putting some constraints on $F(R,G)$ can
restrict these choices (see (\ref{21}) and (\ref{23})). To
elucidate our results we studied GSL in a quasi-de Sitter space
time and a universe with power law expansion.  We showed that in
order that GSL be satisfied in the quasi-de Sitter space time (see
(\ref{24})) a viable $F(R,G)$ must be chosen (see (\ref{28}))
which satisfies the stability condition obtained before in the
literature but not in the framework of GSL. In the case of power
law expansion we obtained an explicit form for $F(R,G)$ and showed
that GSL is satisfied provided that the power of time in the scale
factor be restricted to some special domain specified by the
equation of state parameter of the matter (see (\ref{1000}) and
(\ref{1001})). Despite this, it seems that GSL is violated at
large curvature limit which may be due to quantum effects which
were ignored during our classical computation.

\vspace{1cm} {\bf{Acknowledgments}}

The author would like to thank the Center of Excellence on the
Structure of Matter of the University of Tehran for its supports.


\begin{thebibliography}{99}
\bibitem{acc}
A. G. Riess et al., Astron. J. {\bf 116}, 1009 (1998); S.
Perlmutter et al., Nature (London) {\bf 391}, 51 (1998); P. M.
Garnavich et al., Astrophys. J. {\bf 509}, 74 (1998); S.
Perlmutter et al., Astrophys. J. {\bf 517}, 565 (1999).
\bibitem{infmod}
E. Elizalde, S. Nojiri, S. D. Odintsov, and D. Saez-Gomez,
arXiv:1006.3387v3 [hep-th]; S. Nojiri, and S. D. Odintsov, Phys.
Rev. D {\bf 78}, 046006 (2008); S. Tsujikawa, M. Sami, and R.
Maartens, Phys. Rev. D {\bf 70}, 06325 (2004); S. Nojiri, and S.
D. Odintsov, Phys. Rev. D {\bf 68}, 123512 (2003); I. P. Neupane,
and B. M. N. Carter, JCAP 06, 004 (2006); S. M. Carroll, V.
Duvvuri, M. Trodden, and M. S. Turner, Phys. Rev. D {\bf 70},
043528 (2004); S. M. Carroll, V. Duvvuri, M. Trodden, and M. S.
Turner, Phys.Rev.D {\bf 71}, 063513 (2005); T. P. Sotiriou, and V.
Faraoni, Rev. Mod. Phys. 82, 451 (2010).
\bibitem{string} M. Gasperini and G. Veneziano, Astropart. Phys. 1, 317 (1993).
\bibitem{bousso}R. Bousso, Phys. Rev. D 71, 064024 (2005); S.
Nojiri and S. D. Odintsov, Phys. Rev. D 70, 103522 (2004); Y. S.
Piao, Phys. Rev. D 74, 047301 (2006); B. Guberina, R. Horvat and
H. Nikolic, Phys. Lett. B 636, 80 (2006); B. Wang, Y. Gong and E.
Abdalla, Phys. Rev. D 74, 083520 (2006);  M. D. Pollock and T. P.
Singh, Class. Quantum Grav. 6, 901 (1989); D. Pavon, Class.
Quantum Grav. 7, 487 (1990); R. Brustein, Phys. Rev. Lett. 84,
2072 (2000);  H. M. Sadjadi and M. Jamil, arXiv:1002.3588v1
[gr-qc]; N. Mazumder, and S. Chakraborty, arXiv:1005.5215v1
[gr-qc]; H. M. Sadjadi, Phys. Rev. D {\bf 73}, 063525 (2006).
\bibitem{dav}P. C. W. Davies, Class.
Quantum Grav. 5, 1349 (1988); 4, L255 (1987).
\bibitem{pav}G. Izquierdo, and D, Pavon, Phys. Lett. B. 639, 420
(2006); H. M. Sadjadi, Phys. Lett. B. 645, 108 (2007).
\bibitem{sad} H. M. Sadjadi, Phys. Rev. D {\bf 76}, 104024 (2007);
H. M. Sadjadi, arXiv:1009.1839v1 [gr-qc].
\bibitem{al}M. Alimohammadi, and A. Ghalee, Phys. Rev. D 79, 063006 (2009);
{\bf 80}, 043006 (2009).
\bibitem{app} Y. Gong,  and A.  Wang ,  Phys. Rev. Lett. {\bf 99}, 211301
(2007);  B.  Wang, Y. Gong,  and E. Abdalla,  Phys. Rev.  D {\bf
74},  083520  (2006);  M. Akbar,  and  R.  Cai, Phys. Rev. D {\bf
75}, 084003 (2007);  Y.  Zhang , Z.  Yi ,  T.  Zhang ,  and W.
Liu,  Phys. Rev .D {\bf 77};  023502 (2008 ) A. Frolov,  and L.
Kofman ,  JCAP   05, 009  (2003); M.  Jamil, E.  N. Saridakis, and
M. R. Setare ,  Phys. Rev. D {\bf 81},   023007 (2010); K. Karami,
and S. Ghaffari, Phys. Lett. B {\bf 685}, 115 (2010).
\bibitem{ent}R. M. Wald Phys. Rev. D 48, (1993) 3427; V. Iyer and R. M. Wald,
Phys. Rev. D 50, 846 (1994); T. Jacobson. G. Kang and R. C. Myers,
Phys. Rev. D 49, 6587 (1994); I. Brevik, S. Nojiri, S. D. Odintsov
and L. Vanzo, Phys. Rev. D 70, 043520 (2004); H. Maeda, Phys. Rev.
D 81, 124007 (2010), arXiv:1004.0917 [gr-qc].
\bibitem{stab}G. Cognola, M. Gastaldi, and S. Zerbini,   Int. J. Theor. Phys. {\bf 47}, 898
(2008); A. D. Felice, and S. Tsujikawa, Phys. Lett.B {\bf 675}, 1
(2009); G. Cognola, L. Sebastiani, and S. Zerbini,
arXiv:1006.1586v2 [gr-qc]; A. D. Felice, D. F. Mota, and S.
Tsujikawa, Phys. Rev. D {\bf 81}, 023532  (2010).
\bibitem{end} N. Goheer, R. Goswami, P. K. S. Dunsby and K. Ananda, Phys. Rev.
D 79, 121301 (2009).

\end{thebibliography}
\end{document}